\documentclass[a4paper,twocolumn,showpacs,nofootinbib,floatfix,superscriptaddress,prc]{revtex4}
\usepackage{graphicx}
\usepackage{amsfonts}
\usepackage{amsbsy}
\usepackage{longtable} 
\usepackage{bm} 
\usepackage{hyperref}     
\usepackage{float} 
\renewcommand{\vec}[1]{\mathbf{#1}}
\def\be{\begin{equation}}
\def\ee{\end{equation}}
\def\bea{\begin{eqnarray}}
\def\eea{\end{eqnarray}}

\begin{document}

\title{Comparing the first and second order theories of relativistic dissipative
fluid dynamics using the 1+1 dimensional relativistic flux corrected transport algorithm}

\bigskip

\author{Etele Moln\'ar}
\affiliation{Frankfurt Institute for Advanced Studies, Johann Wolfgang Goethe
Universit\"at, Max-von-Laue-Str. 1, 60438 Frankfurt am Main, Germany}

\begin{abstract}
Focusing on the numerical aspects and accuracy we study a class of bulk
viscosity driven expansion scenarios using the relativistic Navier-Stokes
and truncated Israel-Stewart form of the equations of relativistic
dissipative fluids in 1+1 dimensions.
The numerical calculations of conservation and transport equations
are performed using the numerical framework of flux corrected transport.
We show that the results of the Israel-Stewart causal fluid dynamics are
numerically much more stable and smoother than the results of the standard
relativistic Navier-Stokes equations.
\end{abstract}

\bigskip



\maketitle

\section{Introduction}

The recent discovery of near perfect fluidity of hot QCD matter at the
Relativistic Heavy Ion Collider (RHIC) \cite{RHIC_PF} brought a lot of
attention and interest in modeling the collective phenomena in
relativistic heavy-ion collisions using the relativistic dissipative
fluid dynamical approach.
In contrast to perfect fluid dynamical models, dissipative fluids provide
a more accurate and physically more plausible description incorporating
first and second order corrections compared to perfect fluids.

These higher order corrections are irreversible; thermal conductivity 
and dissipation, related to temperature gradients and
inhomogeneities of the flow field.
A linear relation between the two establishes transport equations,
where the parameters entering these equations are the so-called
transport coefficients, for thermal conductivity $\lambda$, shear
viscosity $\eta$, and bulk viscosity $\zeta$, also referred to as
second viscosity or volume viscosity
\cite{Eckart,Landau_book,Israel_Stewart}.

Recently many studies have specifically investigated the fluid
dynamical description of matter created at RHIC including shear
viscosity, see \cite{Muronga_1,MuroRi,Teaney:2004qa,Heinz_1,BRW,
Chaudhuri_1,Romatschke_2007,Heinz_2} and references therein.
Some of these calculations \cite{Chaudhuri_1,Teaney:2004qa}
made use of the first order theory by Eckart \cite{Eckart}, and Landau
and Lifshitz \cite{Landau_book}, but the main focus was on the second
order causal theory of dissipative fluid dynamics by Israel and Stewart
\cite{Israel_Stewart}, and the theory by \"Ottinger and Grmela \cite{Teaney_1}.

These calculations particularly examined the effect of a small shear
viscosity motivated by the conjectured lower bound from the AdS/CFT
correspondence \cite{Kovtun:2004de,Hirano:2005wx,Csernai:2006zz}.
It was found only recently that, contrary to perturbative QCD estimates
\cite{Arnold_2006}, lattice QCD reveals a large increase of the bulk
viscosity near the critical temperature
\cite{Kharzeev_2007,Karsch_2008,Meyer_2008}. This numerical evidence
motivates studies of the evolution of matter with large viscosity.
It has also been suggested that a large bulk viscosity near $T_c$ may
entirely change the standard picture of adiabatic hadronization
employed so far in hydrodynamical models \cite{TorrieriMishustin}.

In this paper we address the phenomena related to viscous evolution of
matter in 1+1 dimensional systems neglecting the contribution of heat
conduction.
We focus on the numerical implementation of both the 1st-order and 2nd-order
approaches and investigate specific test cases to clarify numerical
aspects and accuracy of the solutions.
This represents an important first step before multi-dimensional
models can be constructed and applied. We also show how the relaxation
equations for the dissipative corrections in the 2nd-order theory can
be solved efficiently and accurately also via the flux corrected
transport algorithm by writing them in the form of continuity
equations with a source.

The paper is organized as follows.
First, we briefly recapitulate and formulate the equations of dissipative
fluid dynamics and the numerical method which will be used to solve the respective equations. 
Afterwards we present and discuss the results in several cases.
To our knowledge major parts of this work including the specific comparisons between the 
1st-order and 2nd-order theories is presented and discussed in detail for the first time.

\section{The equations of dissipative fluid dynamics}

We adopt the standard notation for four-vectors and tensors and use the
natural units $\hbar = c = k = 1$, through this paper.
The upper greek indices denote contravariant while the lower
indices denote covariant objects.
The roman indices or the bold face letters denote three-vectors.

In the Eckart frame, the conserved charge four-current is, $N^{\mu} = n u^{\mu}$,
where $n$ is the local rest frame conserved charge density,
$u^\mu=(\gamma,\gamma \vec{v})$ is the four-flow of matter normalized
to one, $u^{\mu}u_{\mu} = 1$, and the relativistic gamma is,
$\gamma = 1/\sqrt{1 - \vec{v}^2}$.
The dissipative energy-momentum tensor is,
$T^{\mu \nu} = (e + p + \Pi) u^\mu u^\nu - (p + \Pi) g^{\mu\nu} + \pi^{\mu \nu}$,
where, $e = u_{\mu} T^{\mu \nu} u_{\nu}$, is the local rest frame energy
density, the orthogonal projection of the energy-momentum tensor,
$p(e,n) + \Pi = -\frac{1}{3}\Delta_{\mu \nu} T^{\mu \nu}$, denotes the local
equilibrium pressure plus the bulk pressure and
$g^{\mu \nu} \equiv g_{\mu \nu} = \textrm{diag}(1,-1,-1,-1)$,
is the metric of the flat space-time.
The stress tensor,
$\pi^{\mu \nu} = \left[\frac{1}{2} \left( \Delta^{\mu}_{\alpha} \Delta^{\nu}_{\beta} +
\Delta^{\nu}_{\alpha} \Delta^{\mu}_{\beta} \right) - \frac{1}{3}
\Delta^{\mu \nu} \Delta_{\alpha \beta} \right] T^{\alpha \beta}$,
is the symmetric, traceless $\pi^{\mu \nu}g_{\mu \nu} = 0$, and orthogonal to the
flow velocity, $\pi^{\mu \nu} u_{\nu} = 0$, part of the energy-momentum tensor.
The local conservation of charge, energy, and momentum requires that,
\bea\label{baryon}
\partial_{\mu} N^{\mu} &=& 0 \, , \\
\partial_{\mu} T^{\mu \nu} &=& 0 \, ,
\label{energymomentum}
\eea
and the second law of thermodynamics demands that the four-divergence of
the entropy four-current is non-decreasing and positive,
\bea
\partial_{\mu} S^{\mu} &\geq& 0 \, . \label{entropy}
\eea
Here $\partial_\mu = (\partial_t, \partial_i)$ denotes the four-divergence
where $\partial_t \equiv  \partial/\partial t$ is the time-derivative
and $\partial_i \equiv \partial/\partial x_i
= (\partial/\partial x, \partial/\partial y,\partial/\partial z)$ is
the divergence operator.

The explicit form of the conservation equations for charge, energy, and
momentum are,
\bea
\partial_t N^{0} + \partial_i N^{i} &=& 0 \, , \label{cons_n0}\\
\partial_{t} T^{00} + \partial_{i}  T^{0i} &=& 0 \, , \label{cons_e0} \\
\partial_{t} T^{0j} + \partial_{i}  T^{ij} &=& 0 \, , \label{cons_m0}
\label{m_perfect}
\eea
where we defined the conserved charge density, $N^{0}$, charge flux, $N^{i}$,
the total energy density, $T^{00}$, the energy flux density, $T^{0i}$, the
momentum flux density, $T^{i0}$ and the momentum flux density tensor, $T^{ij}$.
These laboratory frame quantities can be expressed in terms of the local rest
frame quantities and velocity as,
\bea
N^{0} &\equiv& n \gamma \, , \label{N0}\\
N^{i} &\equiv& n \gamma v_i \, , \label{Ni} \\
T^{00} &\equiv& (e + p + \Pi )\gamma^2 - (p + \Pi) + \pi^{00}\, , \label{E0} \\
T^{0i} &\equiv& (e + p + \Pi)\gamma^2 v_i  + \pi^{0i}\, , \label{Mi} \\ \nonumber
&=& v_i T^{00} + v_i (p + \Pi) - v_i \pi^{00} + \pi^{0i}\, ,  \\
T^{ij} &\equiv& (e + p + \Pi)\gamma^2 v_iv_j - (p + \Pi) g^{ij} + \pi^{ij}\, , \\ \nonumber
&=& v_i T^{0j} - (p + \Pi)g^{ij} - v_i \pi^{0j} + \pi^{ij} \, .
\eea
The relation between the local rest frame and laboratory frame quantities can
be calculated using the above equations, hence
\bea
n &=& N^{0}\sqrt{1 - v^2} \, , \label{n_LR} \\
e &=& (T^{00} - \pi^{00}) - v_i (T^{0i} - \pi^{0i}) \, , \label{e_LR}
\eea
where the absolute value of the velocity is, $v \equiv |\vec{v}|$.
These local rest frame quantities are needed to calculate the
pressure, $p(e,n)$, from the equation of state (EOS).

The fluid velocity and relativistic gamma can be calculated from eq. (\ref{Mi}), therefore,
\bea
v_i &=& \frac{(T^{0i} - \pi^{0i})}{(T^{00} - \pi^{00}) + P(e,n,\Pi)} \, , \label{vi} \\
\gamma &=& \frac{1}{\sqrt{1 - v^2}} \, , \label{gamma}
\eea
where $P(e,n,\Pi) = p(e,n) + \Pi$, gives the correction to the equilibrium
pressure absorbed in the trace of the energy-momentum tensor.

\subsection{1+1 dimensional expansion}

For simple 1+1 dimensional systems in 1+3 dimensional space-time, where $u^{\mu} = \gamma(1, 0, 0, v_z)$,
the equations of dissipative fluid dynamics reduce to a similar form as in the case
of perfect fluids.
Let us denote the pressure in the longitudinal direction by,
$P_z \equiv p + \Pi + \pi = P(e,n,\Pi) + \pi$,
where $\pi = \pi^{zz}/\gamma^2$ is the local rest frame value of the stress.
Due to construction, the tracelessness property implies that
$\pi^{xx} = \pi^{yy} = -\pi/2$, and
$\pi^{00} = v^{2}_z \gamma^2 \pi$.
The orthogonality relations will further reduce the number of unknowns;
note that, $\pi^{0x} = \pi^{0y} = 0$, $\pi^{0z} \equiv v_z \pi^{zz} = v_z \gamma^2 \pi$,
and all non-diagonal components of the stress tensor vanish, $\pi^{ij}_{i\neq j} = 0$, thus
the only component of the shear tensor we have to propagate is $\pi$ \cite{Muronga_2}.

The conservation equations follow from eqs. (\ref{cons_n0},\ref{cons_e0},\ref{cons_m0}),
\bea
\partial_t N^{0} + \partial_z (v_z N^{0}) &=& 0 \, , \label{cons_n02}\\
\partial_t T^{00} + \partial_z (v_z T^{00}) &=& - \partial_z (v_z P_z) \label{cons_e02}\, , \\
\partial_{t} T^{0z} + \partial_z (v_z T^{0z}) &=& - \partial_z P_z  \label{cons_m02}\, ,
\eea
where the laboratory frame quantities are,
\bea
N^{0} &=& n\gamma \, , \\
T^{00} &=& (e + P_z)\gamma^2 - P_z \, , \\
T^{0z} &=& (e + P_z)\gamma^2 v_z \, .
\eea
The local rest frame variables expressed through the laboratory frame quantities,
the velocity and relativistic gamma are,
\bea
n &=& N^{0} \sqrt{1 - v^2_z} \, , \\
e &=& T^{00}  - v_z T^{0z} \, ,\\
v_z &=& \frac{T^{0z}}{T^{00} + P_z}\, , \\
\gamma &=& \frac{1}{\sqrt{1 - v^2_z}} \, ,
\eea
while the local rest frame effective pressure is,
\bea
P_z &=& p(e,n) + \Pi + \pi \, .
\eea
Here the equilibrium pressure is given by the equation of state, $p = c^2_s e $,
where $c_s$ is the local speed of sound.
The equations and quantities for a perfect fluid are obtained in the limit of
vanishing dissipation corresponding to, $P_z \rightarrow p(e,n)$, while 
the form of conservation equations and the expressions relating
the laboratory frame quantities to the rest frame quantities and the calculation
of the velocity are formally the same as for perfect fluids.

The last two variables that remain to be explicitly defined are the bulk pressure
and the shear.
These can be calculated from  eqs. (\ref{baryon},\ref{energymomentum},\ref{entropy})
using different approaches.
To study the various methods is out of the scope of the current manuscript,
however these theories and recent new phenomenological development aimed to
extend the theory of dissipative fluids sheds light on the open questions
related to the ambiguities on this matter, see for example 
\cite{Liu,Geroch,Gariel,Koide:2006ef,Denicol:2007zz,Van:2007pw,Biro:2008be,Van:2008cy,Osada:2008cn,Osada:2008hr}
and references therein.

In the 1st-order theories of Eckart \cite{Eckart} or Landau and Lifshitz
\cite{Landau_book}, i.e., the relativistic Navier-Stokes equations, 
the entropy four-current is decomposed as,
$S^{\mu} = s u^{\mu} + \beta q^{\mu}$, where $q^{\mu}$ is the heat flux, $s$ is
the local rest frame entropy density and $\beta = 1/T$ is the inverse temperature.
These last two scalar quantities satisfy the fundamental relation of
thermodynamics,  $s = \beta(e+p)$, for matter with no conserved charge.
Hence, in 1st-order theories the only way to satisfy the second law
of thermodynamics, using a linear relationship between the thermodynamic
force and flux, is to choose,
\bea\label{piNS}
\pi_{NS} &\equiv& \pi = - \frac{4}{3}\eta \theta \, , \\
\Pi_{NS} &\equiv& \Pi = - \zeta \theta \, , \label{PiNS}
\eea
where $\eta$ and $\zeta$ are positive coefficients of shear and bulk viscosity
respectively,  while $\theta \equiv \partial_{\mu} u^{\mu}$ is the expansion scalar.
The 1st-order theories (contrary to 2nd-order theories) are known to have
intrinsic problems attributed to the immediate appearance and disappearance of
the thermodynamic flux once the thermodynamic force is turned on or off.
As shown by Hiscock and Lindblom \cite{HL} the linearized version of these
equations propagate perturbations acausally, and even though initially
these might be weak signals they may grow unbounded bringing the system 
out of stable equilibrium.

To remedy some of these problems the 2nd-order theory of Israel and
Stewart was constructed \cite{Israel_Stewart}, similarly to the
non-relativistic theory by M\"uller \cite{Muller}.  This was built
around the assumption that the entropy four-current contains second
order corrections in dissipation due to viscosity (here we disregard
heat conductivity and cross couplings) such that, $S^{\mu} = s u^{\mu}
+ \beta q^{\mu} -(\beta/2) (\beta_0 \Pi^2 + 3\beta_2\pi^2/2) u^{\mu}$,
where $\beta_0$ and $\beta_2$ are thermodynamic coefficients related
to the relaxation times.  Applying the law of positive entropy
production and some algebra leads to the transport equations for the
shear and bulk pressure,
\bea\label{relaxation_pi}
u^{\mu} \partial_{\mu} \pi &=& \frac{1}{\tau_\pi} (\pi_{NS} - \pi) \, , \\
u^{\mu} \partial_{\mu} \Pi &=& \frac{1}{\tau_\Pi} (\Pi_{NS} - \Pi) \, , \label{relaxation_Pi}
\eea
where the relaxation time of bulk viscosity and shear are, $\tau_{\pi} = 2\eta \beta_2$
and $\tau_{\Pi} = \zeta \beta_0$.
The above equations are referred as the truncated Israel-Stewart equations, since
terms involving the divergence of the flow field and thermodynamic coefficients 
have been neglected compared to the equations by Israel and Stewart \cite{Heinz_1}.
However, the current form of the transport equations already captures the essential
features of relaxation phenomena which make the theory causal and stable.

Another crucial difference between first and second order theories is in the
mathematical structure of the equations.
In 1st-order theories the viscous corrections appear linearly proportional
to the divergence of the flow field, therefore they are more sensitive to 
fluctuations and the inhomogeneities in the flow field, see section \ref{results}.
In 2nd-order theory not only the coefficients of viscosity but also the thermodynamic
coefficients need to be specified.
For example the later parameters are know for a relativistic Boltzmann gas
of massive particles \cite{Israel_Stewart}, $\beta_0 = 216 (m\beta)^{-4}/p$, leading to a relaxation time
of, $\tau_{\Pi} =  216(m\beta)^{-4} (\zeta/p)$, which is divergent for a fluid of
massless particles, while the bulk viscosity coefficient is zero in that limit.
The thermodynamic coefficient for the shear viscosity within the same substance is,
$\beta_2 = 3/(4p)$, leading to the relaxation time of, $\tau_{\pi} = (3/2)(\eta/p)$.

In passing we point out a few important facts regarding the bulk viscosity
and its source.
For a long time in the classical non-relativistic Navier-Stokes theory the
purpose of bulk viscosity was controversial \cite{Bulk_1954}.
Even, Eckart in his pioneering work \cite{Eckart} was concerned with fluids
without bulk viscosity.
Israel was the first to show that the bulk viscosity of relativistic matter
may not be unimportant \cite{Israel_1963}.
This turned the attention mainly in cosmology, rendering bulk viscosity as
the only possible form of dissipative phenomena, see for example
\cite{Zimdahl_1996,Maarteens_1996} and references therein for a thorough introduction.
The bulk viscous effects are important in mixtures when the difference in property
between the components becomes substantial.
This might be due to the difference in cooling rates within the same type of
substance or in a mixture between massive and effectively massless particles during
a phase-transition \cite{Maarteens_1997}, (for other possible sources
for bulk viscosity see \cite{Kerstin_2006}).
Therefore in these situations bulk viscosity is used to describe a mixture effectively
as a single fluid with a non-vanishing bulk viscosity coefficient and relaxation time.

It is also important to phenomenologically understand the relation between
viscosity and relaxation time \cite{Landau_book}.
For example, bulk or volume viscosity appears when the system undergoes an isotropic 
expansion or contraction.
If this happens at a relatively fast rate such that the system is unable to follow
the change in volume and restore equilibrium in a short time, means that
the relaxation time of the viscous pressure is long.
On the opposite, if the system equilibrates almost immediately, than the corresponding
relaxation time must be short.
Hence it is also intuitive that large deviations from equilibrium can only be the
consequence of large viscosity, while small departures from equilibrium result
from small viscosity, assuming in both cases that the expansion rate is considerably small.
It is also fundamental that the relaxation time must be shorter than the inverse of
the expansion rate of the system, $\tau_{\pi,\Pi} \ll 1/\theta$, otherwise the system will
never be able to equilibrate and the fluid dynamical approach is unsuitable.

In case of a relativistic Boltzmann gas, recalling the
viscosity coefficients from eqs. (\ref{piNS},\ref{PiNS}), we find that
the relaxation times are
\bea\label{relax_time_pi}
\tau_{\pi} &=& -\frac{d_{\pi}}{\theta}\left(\frac{\pi_{NS}}{p_{eq}} \right)  \, , \\
\tau_{\Pi} &=& -\frac{d_{\Pi}}{\theta}\left(\frac{\Pi_{NS}}{p_{eq}} \right)  \, , \label{relax_time_Pi}
\eea
where $d_{\pi} = 9/8$ and we can only assume similarly to the
relaxation time for the shear, that $d_{\Pi}$ is a dimensionless positive number 
on the order of unity.  
This also follows from the fact that both viscosities give birth to a local dissipative 
pressure, which for small dissipative corrections relax to the Navier-Stokes values.  
In case the dissipative pressure is comparable to the equilibrium pressure, 
the relaxation times become longer than the mean free time between collisions,
thus the fluid dynamical approach may no longer be appropriate.

\section{The numerical scheme}

Here we briefly review the basic principles of the underlying
numerical scheme used in this work.  
The explicit finite difference scheme called sharp and smooth transport algorithm
(SHASTA) \cite{BorisBook_1} is a version of the flux corrected
transport (FCT) algorithm.
Detailed tests, simulations and comparisons to semi-analytical solutions
have been performed with this algorithm in various situations; 
in non-relativistic and relativistic perfect fluid dynamics and magnetohydrodynamics; 
in the last decades \cite{Schneider_1993,SHASTA3d,Dirk_1,Toth_1996,NumVisco2} 
achieving confidence and wide usage.

Before discussing the version of the algorithm in detail, let us
rewrite the conservation equations
(\ref{cons_n02},\ref{cons_e02},\ref{cons_m02}) and transport equations
(\ref{relaxation_pi},\ref{relaxation_Pi}) in conservation form which makes it
possible to treat all equations with the same numerical scheme.
Due to similarity in form and effect in case of 1+1 dimensional expansion scenarios,
we include only one type of viscosity and relaxation equation and refer
to it as bulk viscosity in the following.
Hence,
\bea
\partial_t R + \partial_z (v_z R) &=& 0 \, , \\
\partial_t E + \partial_z (v_z E) &=& - \partial_z (v_z P_z) \, , \\
\partial_{t} M_z + \partial_z (v_z M_z) &=& - \partial_z P_z \, ,
\eea
where $R = N^{0}$, $E = T^{00}$, $M_z = T^{0z}$, and $P_z = p + \pi$ or $P_z = p + \Pi$.
Introducing a common notation, $\tilde{\Phi} = \gamma \Phi$, for the auxiliary variables,
$\tilde{\pi} = \gamma \pi$ and/or $\tilde{\Pi} = \gamma \Pi$, the  relaxation equations
(\ref{relaxation_pi},\ref{relaxation_Pi})
can be rewritten\footnote{Another possibility would be to rewrite the relaxation equation as, 
$\partial_t \Phi + \partial_z (v_z \Phi) = \frac{1}{\tau_\Phi} \left(\Phi_{NS} - \Phi \right) + (\partial_z v_z) \Phi$.} 
in a form similar to the conservation equations,
\bea
\partial_t \tilde{\Phi} +  \partial_z (v_z \tilde{\Phi})
&=& \frac{1}{\tau_\Phi} \left(\Phi_{NS} - \Phi \right) + \theta \Phi \, ,
\eea
where  $\Phi$, $\Phi_{NS}$ and $\tau_\Phi$ commonly denotes the shear pressure 
and its relaxation time and/or the bulk viscous pressure and its relaxation time.

The above conservation and transport equations are of conservation type, written
generally as,
\bea
\partial_t U + \partial_z (v U) = S(t,z) \, ,
\eea
where $U = U(t,z)$ is one of the conserved quantities, $v = v_z$ is the velocity, 
and $S(t,z)$ is the source term.  
The discretised conservative variable defined as an average in cell, $j$, at
coordinate point, $z_j$, at discrete time level, $t^n$, is denoted by $U^n_j$.  
Some of the source terms in our examples contain differential
operations, therefore they are represented as finite (second order)
central differences, i.e., for spatial derivatives $\Delta S_j = (S^{n}_{j+1} - S^{n}_{j-1})/2\Delta z$.

In the SHASTA algorithm the velocity, the local rest frame variables and source terms, 
are computed and updated at half time steps, i.e., in $\Delta t/2$ time intervals.
This requirement ensures second order accuracy in both space and time.
In contrast, the conservative variables, $U$, used to advance the solution 
from time level $n$ to $n+1$, are updated only once at the end of full time steps. 
In a given cell, $j$, this can be summarized formally as, 
\bea \label{U_halfstep}
U^{n+1/2} \sim \tilde{U}^{n} \left( U^{n}, v^{n}, S^{n} \right)\, , \\
U^{n+1} \sim \tilde{U}^{n} \left( U^{n}, v^{n+1/2}, S^{n+1/2} \right) \, .
\eea
In case of the relaxation equation, the source terms contain dynamical information 
on the divergence of the flow field in both space and time. 
Second order accuracy in time can only be calculated in $\Delta t$ time intervals 
(if we use the time-split method), at time levels, $n-1/2,n,n+1/2,n+1, \ldots$, 
where the time derivatives are, 
$\Delta S^{n} \sim (S^{n} - S^{n-1})/\Delta t$, 
$\Delta S^{n+1/2} \sim (S^{n+1/2} - S^{n-1/2})/\Delta t$, etc.
This ensures better accuracy, (however, the difference is rather small), 
than calculating the time derivatives as well as the source terms at full time steps only, i.e., 
only between time levels $n$ and $n+1$.

The difference of primary variables in adjacent cells is denoted by
$\Delta_j \equiv U^{n}_{j+1} - U^{n}_{j}$ or later also by  $\tilde{\Delta}_j = \tilde{U}_{j+1} - \tilde{U}_{j} $.
The explicit SHASTA method \cite{BorisBook_1} at half-step as well as at full-step, first computes 
the so-called transported and diffused quantities,
\bea
\tilde{U}_j &=& \frac{1}{2} \left(Q^2_{+} \Delta_{j} - Q^2_{-} \Delta_{j-1}\right) \\ \nonumber
&+& \left(Q_{+} - Q_{-}\right) U^{n}_{j} + \Delta t \Delta S \, ,
\eea
where
\bea\label{tdiff}
Q_{\pm} = \frac{1/2 \mp \epsilon_j}{1 \pm (\epsilon_{j \pm 1} - \epsilon_{j})} \, , \\
\epsilon_{j} = \lambda v^{n+1/2}_{j}\, ,
\eea
and the Courant number is the ratio of time-step to cell-size, $\lambda = \Delta t/\Delta z$.
A general requirement for any finite difference algorithm is to fulfill the so-called
Courant-Friedrichs-Lewy (CLF) criterion, i.e., $\lambda \leq 1$, related to the stability of
hyperbolic equations, otherwise the numerical solution becomes unconditionally unstable.  
Physically this expresses that, matter must be causally propagated at most $\Delta z = \Delta t$ 
distance into vacuum.  
For SHASTA, $\lambda \leq 1/2$, while in this paper we use a smaller value, $\lambda = 0.4$.  
Here we note that since the numerical algorithms average the transported quantities 
over a cell, part of the matter is acausally propagated over $(1 - \lambda) \Delta z$.
This is a purely numerical artifact called prediffusion.

The time-advanced quantities are calculated removing the numerical 
diffusion by subtracting the so-called antidiffusion fluxes,
$\tilde{A}$, from the transported and diffused quantities such that,
\bea
U^{n+1}_j = \tilde{U}_j - \tilde{A}_j + \tilde{A}_{j-1} \, .
\eea
Here we have defined the flux corrected antidiffusion flux 
\bea\label{antidiff}
\tilde{A}_j &=& \sigma_j \textrm{min}\left[0, \textrm{max}
\left(\sigma_j \tilde{\Delta}_{j+1}, |A_j|,\sigma_j \tilde{\Delta}_{j-1} \right)\right]\, , 
\eea
where the 'phoenical' antidiffusion flux\footnote{The explicit antidiffusion flux \cite{BorisBook_1}, 
$A_j = A_{ad}\tilde{\Delta}_j/8$, leads to somewhat smoother results.} is,
\bea
A_j &=& \frac{A_{ad}}{8} \left[\tilde{\Delta}_j - \frac{1}{8}
\left(\Delta_{j+1} - 2\Delta_{j} + \Delta_{j-1}\right) \right] \, , \\ 
\sigma_j &=& \textrm{sgn}(A_j) \, .
\eea
The so called mask, $A_{ad}$, is introduced to regulate the
amount of anti-diffusion \cite{Book_1}.  
The algorithm tends to produce small wiggles, due to the fact that in the antidiffusion step
one removes to much diffusion, therefore adjusting the mask one
can suppress this artifact leading to a more stable and smoother solutions.  
However the drawback is that, by reducing the antidiffusion we increase the numerical 
diffusion causing larger prediffusion and entropy production even in perfect fluids!
This step is unavoidable in numerical algorithms where due to
discretization the differential equations are truncated already at
leading order and without additional but purely numerical corrections
lead to unstable solutions.  
Within the numerical framework this is called numerical dissipation, or since it 
acts similarly to the physical viscosity it is also called artificial or numerical 
viscosity \cite{Anderson}.

In case we want to model physical viscosity one has to
keep in mind that the there is already a small numerical viscosity in
the algorithm, which has to be estimated (for example by measuring the
entropy production in case of a perfect fluid) and taken into account.
This leads to a total effective viscosity which is larger than the one we explicitly include.
Obviously, things are not as simple, since the numerical viscosity and numerical
diffusion contain linear and non-linear parts \cite{NumVisco2}, and its effect strongly
depends on the grid size, initial condition and flux limiters we use.
Therefore it is a question of numerical analysis and extensive testing 
to reveal the effect of numerical viscosity. 
Some can be found in the original or related publications of the numerical schemes.

It is important to remember that SHASTA is a low implicit viscosity algorithm,
and conserves energy (and momentum) up to 5-digits, but produces entropy in the case of a perfect fluid
roughly between $0.5\% - 5\%$, depending on the initial setup, antidiffuison flux, mask coefficient and 
physical situation.
The lower value was found in the studies we are going to show in the next section using a mask of $A_{ad} = 0.8$,
while the error is less than $0.2 \%$ using the standard value, $A_{ad} = 1$, 
after 200 time-steps.
The large entropy production was found in the case of a 3D grid with the same proportions,
cell size, number of time-steps and reduced mask coefficient in case of 1+3 dimensional
expansion into vacuum of an initially constant energy sphere.

To determine the bulk viscous pressure one also has to calculate the
expansion scalar, $\theta(t,z)$.  
One possibility is to take the standard form, used in this work,
using a second order accurate central difference formula, 
$\theta \equiv \partial_t \gamma + \partial_z (v_z \gamma) 
= \gamma^3 (v_z\partial_t v_z + \partial_z v_z)$. 
The other form can be expressed from the
conservation of energy, $u_{\nu} \partial_{\mu} T^{\mu \nu} = 0$, or
in case we also have conserved charge, from the continuity equation,
$\partial_{\mu} N^{\mu} = 0$, leading to, $\theta = - \gamma
\left(\partial_t e + v_z \partial_z e \right)/(e + p + \Pi + \pi) =
-\gamma\left(\partial_t n + v_z \partial_z n \right)/n$.

The numerical differentiation of the velocity field introduces obvious
numerical problems which we need to address. In particular, finite
differences of the velocity field in adjacent cells, usually fluctuate
due to numerical noise. Since we solve a set of non-linear coupled
partial differential equations, these may become uncontrollable. To
make the numerical expansion rate smoother we found that an additional
five-point stencil smoothing is
necessary\footnote{In a loose sense
this coarse-graining of the expansion rate may be viewed as
providing a ''mass'' to fluctuations with wave-length on the order
of the grid spacing.}.
However, the maximum number of neighboring cells to include is restricted
by the Courant number, otherwise we acausally propagate information into
the neighboring cells.
In our example the maximal number of these cells are two to the right and two
to the left, hence the five-point stencil.

In the 1st- and 2nd-order theories the dissipative pressure, $|\Pi|
\leq p_{eq}$, must be smaller than the equilibrium pressure. 
If the correction to the equilibrium pressure is small, the system will 
continue to expand with a lower effective pressure but the overall behavior
should not change considerably from the perfect fluid limit. 
However, at different parts of the system the local expansion rate may become 
very large (for example, in the transition region to vacuum) and generate 
large dissipative corrections.  
This threshold is given by the equilibrium pressure locally, at least in 
the 1st-order theory.
Hence, even though the physical situation may encounter larger values of
the bulk pressure, we choose to keep this maximum.  
The upper bound imposed on the bulk pressure leads to an
upper bound for the local expansion rate, $\theta_{max} = p_{eq}/\zeta$.  
In other words, the Navier-Stokes bulk pressure is
defined to be $\Pi_{NS} = -\zeta \theta$ for $\theta < \theta_{max}$
and $\Pi^{max}_{NS} = -\zeta \theta_{max}$ for $\theta \ge \theta_{max}$. 
In the latter region the total pressure vanishes and the acceleration 
therefore stalls.

We keep the above convention also for 2nd-order theory so that for
very short relaxation times we exactly approach the Navier-Stokes limit.  
It is also important to mention that since we solve the
relaxation equation in conservation form, the expansion rate appears
explicitly in the truncated equations as well, similarly as in the
full Israel-Stewart equations.  Hence it is obviously necessary to
find an upper bound for terms containing the expansion rate otherwise
those terms may grow unbounded and destabilize the solutions.

\section{Results and discussion}\label{results}

In all numerical calculations, we have fixed the following parameters.
The Courant number, $\lambda = 0.4$, and the cell size is $dz = 0.2$
fm, therefore $dt = 0.08$ fm/c.  The grid contains 240 cells, while
the total number of time steps is, $n = 200$, which corresponds to
$\Delta t = 16$ fm/c expansion time.  The amount of anti-diffusion is
reduced by $20\%$, i.e., $A_{ad} = 0.8$, which leads to some
prediffusion but returns smoother profiles.  
The thermodynamic quantities are given by the Stefan-Boltzmann relations 
where the degeneracy of massless particles is $g=16$. 
In case of dissipative fluids, the bulk viscosity coefficient to entropy density is 
constant with values corresponding to small, $\zeta/s = 0.2$, 
and to large, $\zeta/s = 1$, ratios.

In most situations the initial expansion rate is unknown, therefore
the dissipative corrections are neglected at start.  
This may only last for one time-step, since after that the time-derivatives 
can already be calculated and dissipative corrections added.  
In particular cases such as the Bjorken scaling solution, the expansion
rate can be inferred from the geometry, therefore it does not pose a
problem.  We will return to this issue in Sect. \ref{bjsection}.

The relaxation time is given similarly to eq. (\ref{relax_time_Pi}),
therefore $\tau_{\Pi} = -\frac{1}{\theta}\left(\frac{\Pi_{NS}}{p_{eq}}
\right) $.  We have checked the asymptotic limits of the
relaxation equations.  In case the relaxation time is small,
$\tau_{\Pi} \approx dt$, the effect of bulk viscosity is immediately
felt by the system, therefore the system behaves as in case of 1st-order theories.
For very large relaxation times, i.e., larger than
the lifetime of the system, and small initial dissipation, the effect of
viscosity is exponentially suppressed and the solution approaches 
the perfect fluid limit.

\subsection{Expansion into vacuum of perfect fluid}\label{riemann}

In case of perfect fluids one of the analytical solutions which
relates the thermodynamic properties of matter and the type of fluid
dynamical solution is the 1+1 dimensional expansion into vacuum.  
This is a special case of the relativistic Riemann problem describing 
one-dimensional time-dependent flow. 
The initial conditions are such that initially at $t=0$ half
of the space, $z\leq0$, is filled uniformly with fluid at rest,
$v(z,0) = 0$, with energy density, $e(z,0)=e_0$, while the positive
half $z>0$ is (empty) filled with vacuum.

\begin{figure}[!htb]
\centering
\includegraphics[width=7.5cm, height = 8.5cm]{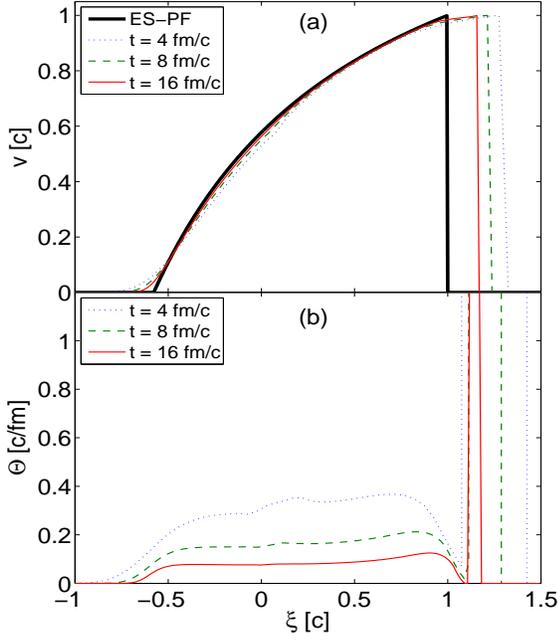}
\caption{The exact solution in case of a perfect fluid (thick line) and numerical solutions (thin dotted line at $t=4$ fm/c, thin dashed line at,
$t=8$ fm/c, and thin line at $t=16$ fm/c) as function of the similarity variable $\xi(z,t)$.
(a) the flow velocity $v(\xi)$; (b) the expansion scalar $\theta(\xi)$.}
\label{fig01}
\end{figure}

One can show that for thermodynamically normal matter, 
e.g., a massless ideal gas with the EOS, $p(e) = c^2_s e$, where $c^2_s=1/3$
is the speed of sound squared, the stable solution to the fluid
dynamical equations is a simple rarefaction wave.  The rarefaction
wave is a wave where the energy density decreases in the direction of
propagation, but the profile of the flow does not change with time 
as a function of the similarity variable,
\bea \label{simvariable}
\xi \equiv \frac{z}{t} = \frac{v(e) - c_s}{1 - v(e)c_s} \, .
\eea
Here we recall the analytic results for a perfect fluid in the forward
light-cone \cite{Dirk_1}.  The energy density as a function of the
similarity variable for, $-1 \leq \xi \leq -c_s$, is constant,
\bea
e(\xi) &=& c^{-2}_s p_0 \, ,
\eea
while in the region $-c_s \le \xi
\leq 1$ the matter starts to rarefy and the energy density decreases
as
\bea e(\xi) &=& c^{-2}_s p_0
\left(\frac{1 - c_s}{1 + c_s}\frac{1 - \xi}{1 + \xi} \right)^{(1 + c^2_s)/(2c_s)} \, .
\eea
The temperature can be inferred from standard thermodynamical
relations and the EOS, leading to
\bea \label{a_temp}
T(\xi) &=& T_0
\left(\frac{e(\xi)}{e_0}\right)^{c^2_s/(1 + c^2_s)} \, .
\eea
We also compare how well the fluid flow is reproduced by the numerical
calculation, where the analytical solution as a function of energy density is,
\bea v(e) = \frac{1 - (e/e_0)^{2c_s/(1+c^2_s)}}{1 + (e/e_0)^{2c_s/(1+c^2_s)}} \, .
\eea
The velocity can be given as a function of the similarity variable as
well from eq. (\ref{simvariable}), this is plotted in Fig. \ref{fig01}a.
The results for the expansion scalar calculated numerically are shown in
Fig. \ref{fig01}b.

\begin{figure}[!htb]
\centering
\includegraphics[width=7.5cm, height = 8.5cm]{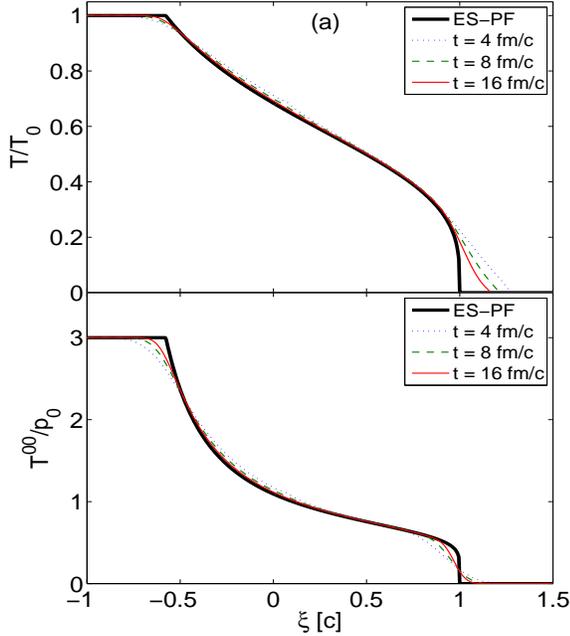}
\caption{(a) the temperature normalized to the initial temperature, $T(\xi)/T_0$;
(b) the laboratory frame energy density normalized to the initial pressure $T^{00}(\xi)/p_0$;
both as function of the similarity variable $\xi(z,t)$.
The paramteres are the same as in Fig. \ref{fig01}.
}
\label{fig02}
\end{figure}

Fig. \ref{fig02}a shows the temperature normalized by the initial
temperature, from eq. (\ref{a_temp}), while Fig. \ref{fig02}b, shows 
the lab frame energy density normalized by the initial pressure 
as a function of the similarity variable.
The thick line shows the exact solution for a perfect fluid (ES-PF), while
the numerical solutions also for a perfect fluid are at $t=4$ fm/c with
thin dotted line, at $t=8$ fm/c with thin dashed line, and at $t=16$ fm/c
with full thin line.

Both figures compare the analytical solutions to the numerical solutions,
to pinpoint how well the underlying numerical method reproduces the ES-PF.
We see that the numerical solutions asymptotically approach the
exact result, while also reduce the numerical prediffusion into vacuum.
This is due to the fact that the initially sharp discontinuity smears out
as the rarefaction wave covers an increasing number of cells.
Due to prediffusion and coarse-graining, the
expansion rate it is not zero for $\xi > 1$ as can be seen in both figures; 
this problem mostly affects the acausally propagated low density matter.
For a perfect fluid the numerical results are smooth and very nicely reproduce
(especially at later times) the exact results.
The larger deviations from the ES-PF around the boundary to vacuum is due to a somewhat
large prediffusion caused by the reduced antidiffusion, while the deviations around 
$-c_s$ are due to a larger numerical diffusion.
A more thorough study of the expansion into vacuum of a perfect fluid can be found in Ref. \cite{Dirk_1}.

\subsection{Expansion into vacuum with small and large dissipation}\label{sl_diss}

Here we analyze and study the behavior of dissipative fluids corresponding to the 1st-order 
and 2nd-order theories, and plot the numerical results next to the exact solution in case of a perfect fluid.
In all figures, unless stated otherwise, all quantities are plotted as a function of the similarity variable; 
the ES-PF is plotted with dotted line, while the numerical solutions at $t=4$ fm/c with thin dashed line 
and at $t=16$ fm/c with continuous line.
The upper bound for the bulk pressure is, $\Pi^{max} \equiv p_{eq}$, and the
relaxation time is,  $\tau_{\Pi} = -\frac{1}{\theta}\left(\frac{\Pi_{NS}}{p_{eq}} \right) $fm/c.
The ratio of viscosity over entropy density corresponds to small $\zeta/s = 0.2$ or large $\zeta/s = 1.0$ values.

\begin{figure}[!htb]
\centering
\includegraphics[width=7.5cm, height = 8.5cm]{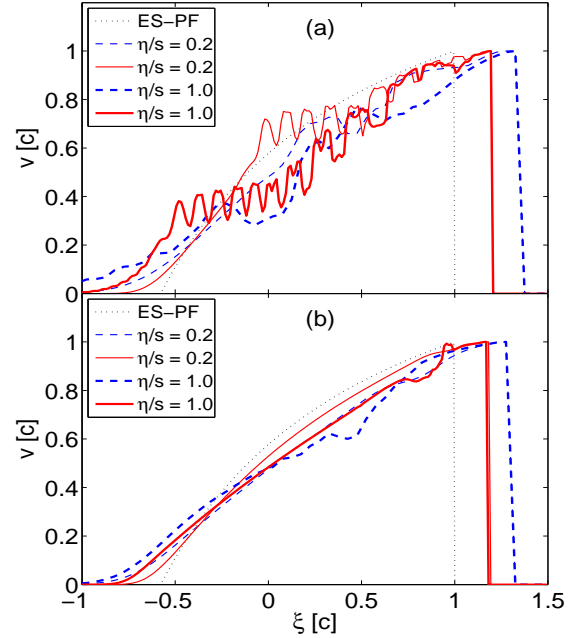}
\caption{The velocity profile for a perfect fluid ES-PF compared to dissipative fluids with
$\zeta/s = 0.2$ (thin) and $\zeta/s = 1.0$ (thick), as numerically calculated and plotted 
at $t = 4$ fm/c (dashed) and $t = 16$ fm/c (full);
(a) 1st-order theory;
(b) 2nd-order theory.
}
\label{fig03}
\end{figure}

Fig. \ref{fig03}a shows the velocity profile calculated from 1st-order theory,
while Fig. \ref{fig03}b from 2nd-order theory.
We see that for 1st-order theory the velocity is fluctuating with an increasing frequency in time.
Initially the amplitude and the wavelength of the fluctuations is large, however as the expansion
proceeds these large fluctuations are damped and become smaller amplitude and smaller wavelength oscillations.
This is partially due to that, that the initial discontinuity smears out in time and the problem is
resolved on a larger grid.
However, the damping is also due to the non-linear antidiffusion term (\ref{antidiff}) in 
the numerical algorithm.
In other words this is numerical viscosity, which always acts to smooth the fluxes,
contrary to the numerical dispersion which acts exactly the opposite and produces ripples in the results.
It should also be obvious that working on a larger grid with the same cell size would not improve the results!

For bigger viscosity the numerical fluctuations become larger which is clearly a sign that 
the method fails and the numerical errors become uncontrollable.
The results for 2nd-order theory are much smoother and show that there is
a relevant change in the velocity profile, which is an outcome of the large dissipative pressure.
This interesting phenomena appears near the edge of the matter where due to a large bulk pressure
contribution the effective pressure essentially vanishes, therefore that part of the system
stops to accelerate but due to inertia it keeps moving forward with constant speed, 
hence forming a constant velocity plateau. 
Note that, the small wiggle in the velocity profile for large dissipation, visible for example 
on Fig. \ref{fig03}b, is an artifact of the phoenical antidiffusion flux.

\begin{figure}[!htb]
\centering
\includegraphics[width=7.5cm, height = 8.5cm]{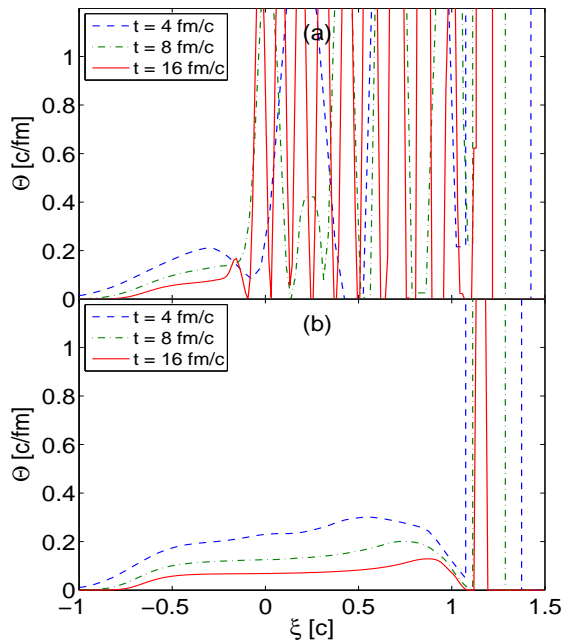}
\caption{The calculation of the expansion scalar at different time-steps in case of 
a dissipative fluid with $\zeta/s = 0.2$.
(a) 1st-order theory;
(b) 2nd-order theory.
}
\label{fig04}
\end{figure}

\begin{figure}[!htb]
\centering
\includegraphics[width=7.5cm, height = 8.5cm]{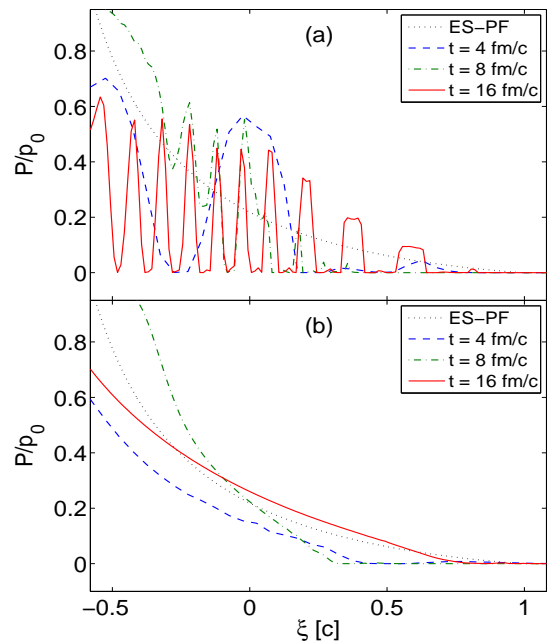}
\caption{The effective pressure normalized by the initial pressure with dissipation proportional to $\zeta/s = 1$.
(a) 1st-order theory;
(b) 2nd-order theory.
}
\label{fig05}
\end{figure}

Figs. \ref{fig04}a and \ref{fig04}b shows the expansion scalar calculated numerically at different
time-steps for both theories.
We can see that in case of 1st-order theory, the expansion scalar is highly fluctuating,
while in 2nd-order theory it is much smoother.
Both calculations reflect the space-time inhomogeneity of the flow field amplified by the velocity.

For a better understanding we have plotted the effective pressure as a function of the similarity
variable in Fig. \ref{fig05}, for both theories.
One can see that for early times the expansion rate and therefore the bulk
viscous pressure is largest.
When the effective pressure drops (to zero), the acceleration of matter is reduced 
(the matter flows with constant velocity).
However, the expansion rate will also decrease later, reducing the viscous pressure and therefore
the velocity will continue to increase.
The important thing to remember is that the speed of sound decreases due to dissipation 
and the effective rarefaction speed \cite{Mornas:1994yp} can be given as, $c^2_{eff} \sim (p + \Pi)/e$.

It is also interesting to remark that even though the bulk viscosity is large at places,
the effective pressure (and therefore the equilibrium pressure) may be larger than in the
case of perfect fluid, due to the fact the thermodynamic state of the system is influenced by
entropy production and slower cooling.
This is why the lab frame energy density decreases slower, see Fig. \ref{fig07}.

\begin{figure}[!htb]
\centering
\includegraphics[width=7.5cm, height = 8.5cm]{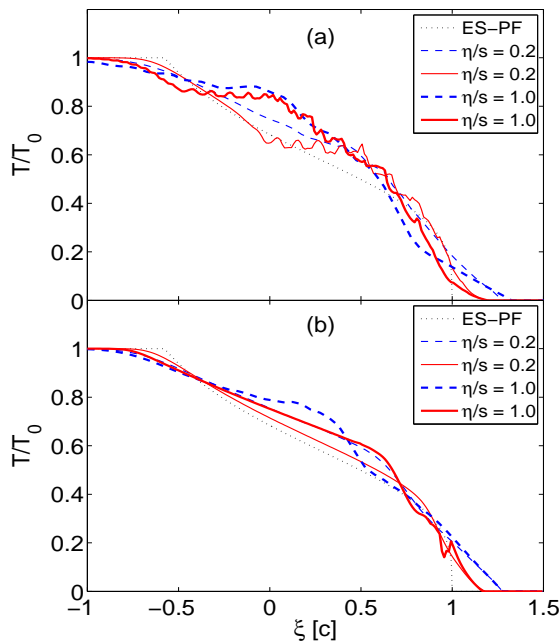}
\caption{The temperature normalized by the initial temperature.
(a) 1st-order theory;
(b) 2nd-order theory.
The parameters are the same as in Fig. \ref{fig03}.
}
\label{fig06}
\end{figure}

Fig. \ref{fig06} shows the temperature profile plotted at different time-steps for both theories.
As in the case of velocity the presence of viscosity
is observable in the overall reduced cooling of matter.
The increase in the expansion rate increases the dissipation which in turn slows the expansion,
thus the matter cools at a slower rate.
On the other hand in 1st-order theories, the visibility  of this effect is much more reduced due 
to numerical problems.

\begin{figure}[!htb]
\centering
\includegraphics[width=7.5cm, height = 8.5cm]{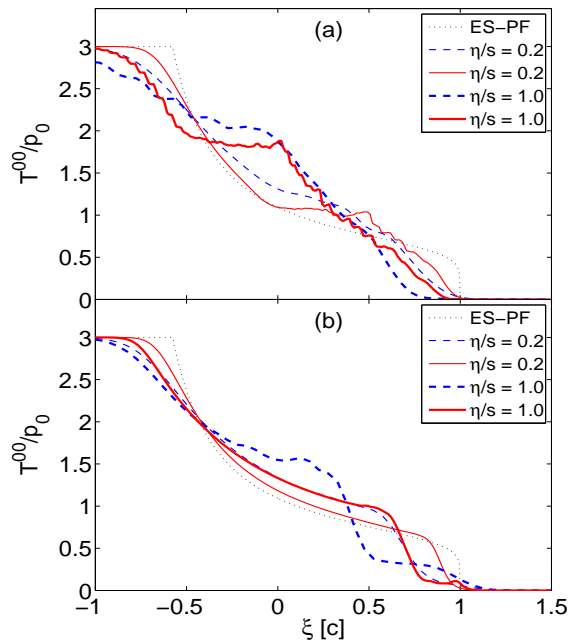}
\caption{The laboratory frame energy density normalized to the initial pressure.
(a) 1st-order theory;
(b) 2nd-order theory.
The parameters are the same as in Fig. \ref{fig03}.
}
\label{fig07}
\end{figure}

The laboratory frame energy density normalized to the initial pressure is presented in Fig. \ref{fig07}.
Since this quantity is proportional to the fourth root of the temperature divided by
the initial temperature, it is much less affected by the fluctuations and prediffusion in both cases.
Based on the previous arguments the deviation from the ES-PF are noticeable for larger values
of the similarity variable, where the effect of dissipation is most pronounced.
This plot also confirms that the effective pressure drops to zero, however since 
the matter has a finite energy density, temperature, and velocity, the 
laboratory energy density is not zero at those places.
As soon as the expansion rate decreases, the finite albeit small effective pressure will continue
to expand the matter into vacuum.
Further comparisons between the 1st-order and 2nd-order theories can be found in the Appendix.

\subsection{Expansion into vacuum with a soft EOS}

To further investigate the behavior of matter with large viscosity we have used a relatively soft EOS,
where the speed of sound squared is $c^2_s = 1/15$, while keeping the other parameters intact.
Using a soft equation of state reduces the pressure and the pressure gradients in the system, while
the relaxation time increases, $\tau_{\Pi} \sim (1 + c^2_s)/(c^2_s \ T)$, inversely with
local speed of sound and temperature.

\begin{figure}[!htb]
\centering
\includegraphics[width=7.5cm, height = 8.5cm]{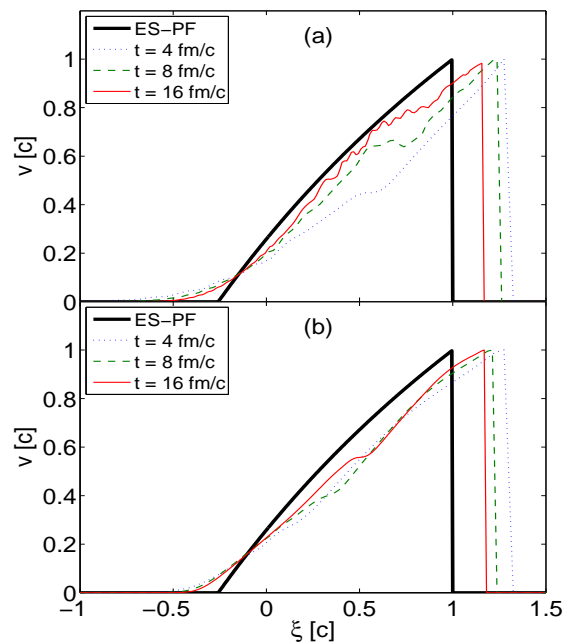}
\caption{The velocity profile for a perfect fluid, (ES-PF), compared to  a dissipative fluid with
$\zeta/s = 1$,   using a soft EOS.
(a) 1st-order theory;
(b) 2nd-order theory.
}
\label{fig08}
\end{figure}

\begin{figure}[!htb]
\centering
\includegraphics[width=7.5cm, height = 8.5cm]{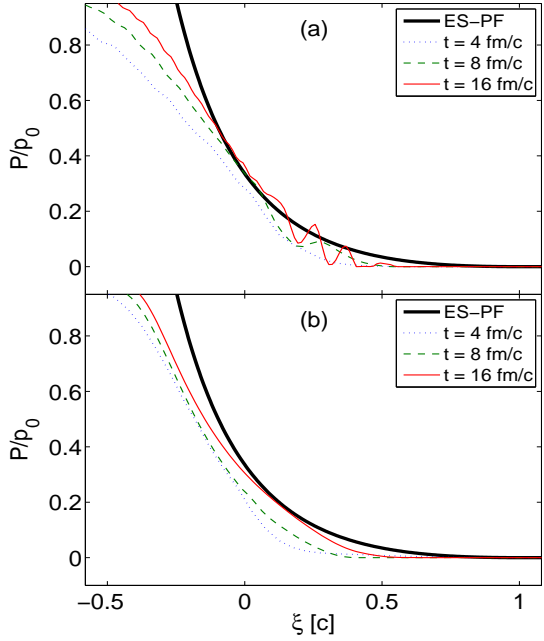}
\caption{The effective pressure normalized by the initial pressure for a soft EOS.
(a) 1st-order theory;
(b) 2nd-order theory.
}
\label{fig09}
\end{figure}

Fig. \ref{fig08}a shows the velocity profile for 1st-order theory
while Fig. \ref{fig08}b for 2nd-order theory, both in case of a soft EOS.  
In comparison to Fig. \ref{fig03}, the velocity profiles
are much improved and the constant velocity part of velocity profile is
clearly visible even from the 1st-order theory.  
It is clear that the algorithm works much better, producing overall smooth results. 
We have also checked our algorithm with a hard EOS, $e=p$, which proved that 
overshoots and oscillations become enhanced compared to the standard 
case ($c_s^2=1/3$) and the results became less reliable.

Fig. \ref{fig09}, shows the effective pressure normalized by the initial pressure, 
similar to Fig. \ref{fig05}, but with a soft EOS.  
Once again the results are good, especially in the case of 2nd-order theory.  
This is due to fact that a softer EOS not only reduces the thermodynamic pressure 
compared to the standard EOS but also decreases the dissipative pressure.  
It is also important to understand that in this
case the dissipative effects act on an overall wider scale as function of $\xi$,
but even though the effect of dissipation is immediately added, the
results are much smoother because the dissipative pressure is also reduced.

\subsection{Expansion into vacuum of a matter with temperature dependent bulk viscosity}

A interesting and relevant question to study, is the expansion of
dissipative matter with a temperature dependent bulk viscosity.  
To model this property, we assumed that bulk viscosity acts in
the close vicinity of a specific or critical temperature, 
$T = T_c (1 \pm 0.02)$, otherwise it is zero.  
Thus the bulk viscosity coefficient is,
\bea
\zeta = \zeta_0 \Theta(T - 1.02 T_c) \left(1 - \Theta(T - 0.98T_c) \right) \, .
\eea
Here the choice of critical temperature is, $T_c = 2T_0/3$, where
$T_0$ is the initial temperature, and $\zeta_0 = s$ is the bulk
viscosity coefficient.  
The calculations are done for 2nd order theory including the above temperature 
dependent bulk viscosity.

\begin{figure}[!htb]
\centering
\includegraphics[width=7.5cm, height = 8.5cm]{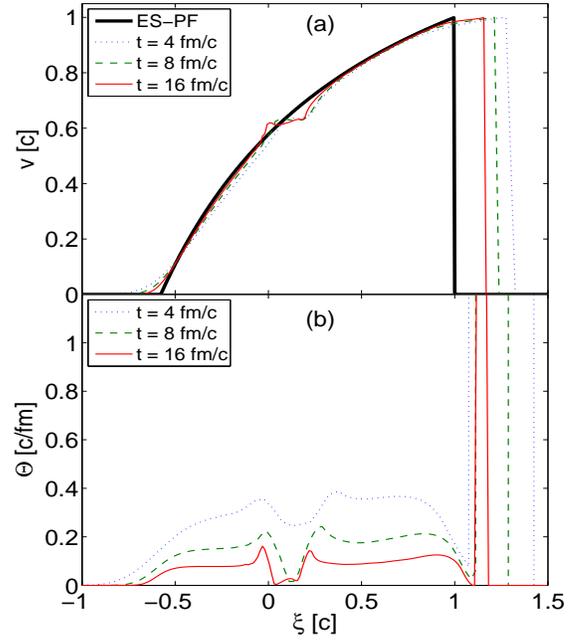}
\caption{
The velocity profile (a), and the numerical calculation of the expansion scalar (b),
using a temperature dependent bulk viscosity for 2nd-order theory.
}
\label{fig10}
\end{figure}

\begin{figure}[!htb]
\centering
\includegraphics[width=7.5cm, height = 8.5cm]{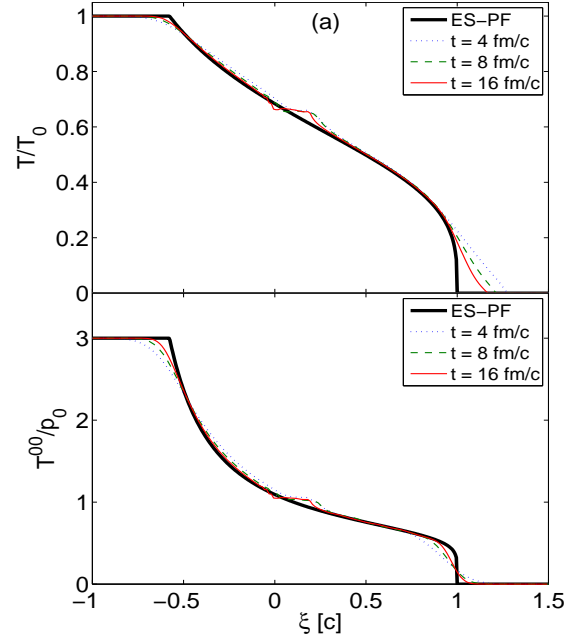}
\caption{The parameters are the same as in Fig. \ref{fig10}.
(a) the temperature normalized to the initial temperature, $T(\xi)/T_0$;
(b) the laboratory frame energy density normalized to the initial pressure $T^{00}(\xi)/p_0$.}
\label{fig11}
\end{figure}

Figs. \ref{fig10}a and \ref{fig10}b shows the velocity of the matter
and the expansion rate as a function of the similarity variable.  For
the same setup, in Figs. \ref{fig11}a and \ref{fig11}b the temperature
normalized by the initial temperature and the laboratory frame energy
density normalized by the initial pressure is shown.

When the temperature falls in the respective regime, the viscosity over entropy ratio 
rises suddenly, to $\zeta_0/s = 1$, while otherwise the dissipation is switched off.  
In our case this manifests itself as almost constant velocity temperature, pressure, 
and energy density plateau, located roughly between $0 < \xi < 0.2$.  
Because the effective pressure decreased suddenly, the matter is slowed down considerably, 
until the matter cools below the predefined temperature (although much slower), 
then the system will suddenly find local thermal equilibrium and continue to accelerate.
This is apparent in all plots.
This type of studies may be relevant in case of a phase-transitions
where the temperature dependent viscosity modeling becomes necessary \cite{Torrieri:2008ip}.

\subsection{The Bjorken solutions for perfect and dissipative fluids}\label{bjsection}

In this section we test how well the numerical calculations reproduce
one-dimensional dissipative scaling flow. 
The relaxation time was kept constant, $\tau_{\Pi} = 1$~fm/c, 
which does not effect the final outcome qualitatively.

We first recall the 1+1 dimensional Bjorken scaling solution for perfect \cite{Bjorken:1982qr} and
dissipative fluids \cite{Muronga_1}.  The equations follow from the conservation law
$\partial_{\mu} T^{\mu \nu} = 0$, and the second law of
thermodynamics, under the assumption that the matter expands
longitudinally with a flow velocity $v = z/t$ in a boost invariant manner.
To transform the partial differential equations into simple differential
equations (using the assumption of boost invariance), one carries
out a coordinate transformation from $(t,z)$ to $(\tau,\eta)$, where
$\tau = \sqrt{t^2 - z^2}$ is the proper time and
$\eta = \frac{1}{2} \log \left[(t + z)/(t - z)\right]$ is the space-time rapidity.
Therefore, the truncated Israel-Stewart equations for the energy and bulk pressure are,
\bea \label{e_BJ} 
\frac{d e}{d\tau} 
&=& -\frac{1}{\tau}\left(e + p + \Pi\right) \, , \label{e_Bj}\\
\frac{d\Pi}{d\tau} &=&
\frac{1}{\tau_{\pi}} \left( \Pi_{NS} - \Pi \right) \, , \label{Pi_BJ}
\eea
where $\Pi_{NS} = -\zeta/\tau$, and the effective pressure satisfies, $d(p+\Pi)/d\eta = 0$.
The equations of perfect fluid dynamics are obtained 
when the dissipative bulk pressure is zero, $\Pi(\tau) = 0$.  
The relativistic Navier-Stokes equation is given by eq. (\ref{e_BJ})
alone, since bulk pressure $\Pi(\tau) = \Pi_{NS}(\tau)$ is given
algebraically.

In the Bjorken picture the system is infinitely elongated in rapidity.
Since our SHASTA code is written in standard space-time coordinates
$(t,z)$, we have to determine initial values on a $t=t_0$ surface and
the fluid must have a finite length (due to the finite grid), $-z_0
\leq z \leq z_0$.  This is done as follows.

We first solve eqs. (\ref{e_BJ},\ref{Pi_BJ}) using a 4th-order
Runge-Kutta solver for all times $\tau\ge\tau_0=1$~fm/c. Together with
the assumption of boost invariance, this determines the hydrodynamic
fields in the entire forward light-cone. We then repeat the solution
with the SHASTA partial differential equation solver with initial
conditions at $t_0 = z_0 + \Delta z=6$~fm set by the Runge-Kutta solution. We
note that in the SHASTA solution the system has a boundary since the
velocity at $z_0$ is close to but slightly less than the velocity of
light ($z_0$ is smaller than $t_0$ by one grid spacing).

\begin{figure}[!htb]
\centering
\includegraphics[width=7.5cm, height = 8.5cm]{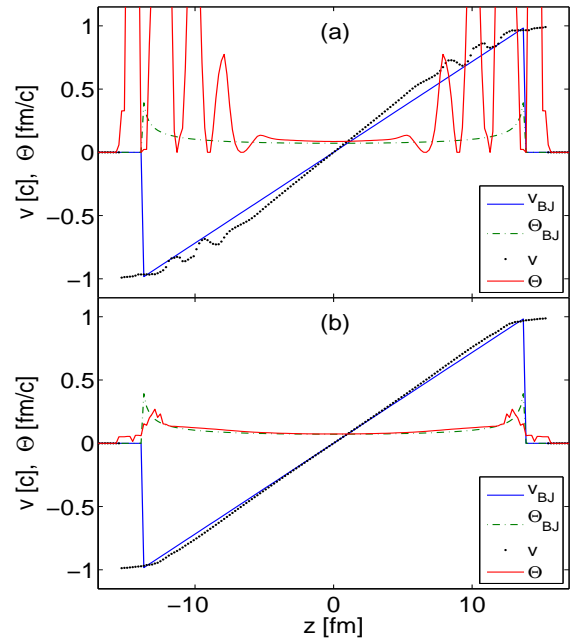}
\caption{The exact velocity v$_{BJ}$ (full line) and expansion rate $\Theta_{BJ}$ (dash dotted line),
compared to the numerically calculated velocity v (dotted line) and expansion rate $\Theta$
(full line), for a dissipative fluid with $\zeta/s = 0.2$ after $\Delta t = 8$ fm/c evolution.
(a) 1st-order theory;
(b) 2nd-order theory with $\tau_{\Pi} = 1$ fm/c constant relaxation time.
In both cases the initial value for the bulk pressure is given by the Navier-Stokes value.
}
\label{fig12}
\end{figure}

In the 1st-order theory the initial value for the bulk pressure is
$\Pi_{NS} = -\zeta_0/\tau_0$, which can be limited by the equilibrium
pressure, i.e., the dissipative pressure should not be larger than the
initial equilibrium pressure otherwise the system becomes unstable.
In the 2nd-order theory, we take the initial value of the bulk
pressure to be the same as in the 1st-order theory. Using this initial
condition allows for a direct comparison since both theories start
from the same initial values. Moreover, on physical grounds, if
$\tau_0$ is to be interpreted as the onset of hydrodynamic behavior
(thermalization time) than the Reynolds number at $\tau_0$ should be
stationary, i.e., it should neither grow nor decrease; this implies
that the initial value for the bulk pressure should be close to that
given by the 1st-order approach~\cite{DMN}.

\begin{figure}[!htb]
\centering
\includegraphics[width=7.5cm, height = 8.5cm]{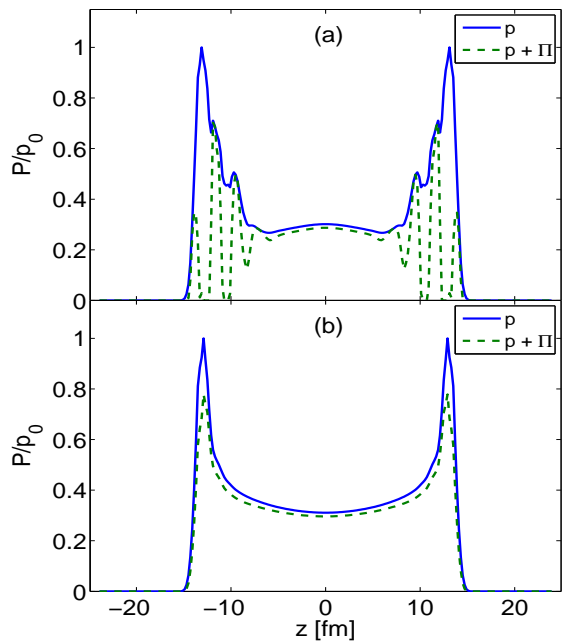}
\caption{The effective pressure normalized by the initial pressure.
(a) 1st-order theory;
(b) 2nd-order theory.
The parameters are the same as in Fig. \ref{fig12}}
\label{fig13}
\end{figure}


We can see in Fig. \ref{fig12} that the flow velocity and expansion
rate are fairly well reproduced for the 2nd-order theory while the
results for the 1st-order theory show large numerical artifacts
already at a few fm distance from the center. This is correlated with
the coarse graining since the results improve on finer
grids\footnote{In view of forthcoming applications to
three-dimensional problems we only consider grids with at most a few hundred grid cells.}. 
We can also observe the large prediffusion into vacuum due to reduced antidiffusion 
similarly to the Riemann wave discussed above. 
The fluctuations in the velocity are also visible in the expansion rate and in 
the pressure shown in Fig. \ref{fig13}.

\begin{figure}[!htb]
\centering
\includegraphics[width=7.5cm, height = 8.5cm]{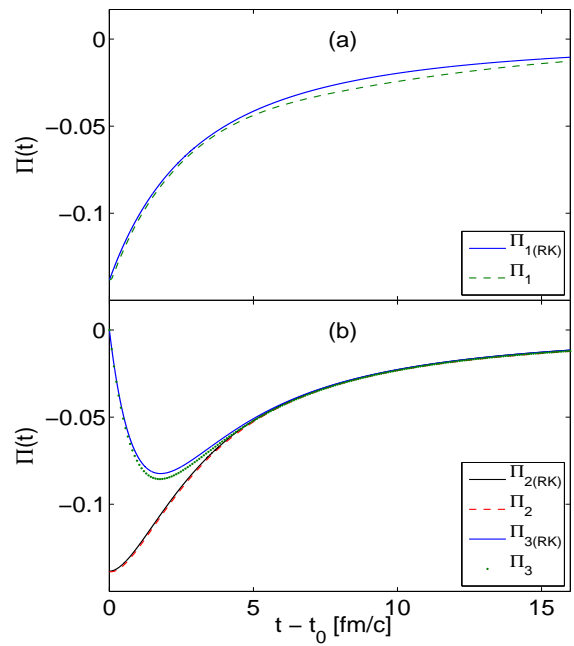}
\caption{The evolution of the bulk pressure in a central cell.
(a) 1st-order theory;
(b) 2nd-order theory.
The initial condition for the bulk pressure is given by the Navier-Stokes value for,
$\Pi_{1}$, $\Pi_{2}$, $\Pi_{1(RK)}$ and $\Pi_{2(RK)}$.
The evolution of the initially equilibrated system are for, $\Pi_3$ and $\Pi_{3(RK)}$.
}
\label{fig14}
\end{figure}

We have also tested how well the algorithm solves the relaxation
equation for the bulk pressure.  This is important to know, since the
SHASTA was specially designed to solve conservation equations.
We have plotted the evolution of the bulk pressure in the central cell,
which has a velocity $v \approx 0$, next to the results of a standard 
fourth-order Runge-Kutta solver.  
The comparison in case of 1st-order theory is given in Fig.~\ref{fig14}a.  
Here $\Pi_{1(RK)}$ with full line is the Runge-Kutta solver, while the 
result of SHASTA is $\Pi_{1}$, with dashed line. 
The curves match reasonably well, deviations are due to
the overestimate of the local expansion rate.

For 2nd-order theories our standard choice for the initial value of
the bulk pressure is given by the Navier-Stokes value; this
corresponds to $\Pi_{2(RK)}$ and $\Pi_{2}$ in Fig.~\ref{fig14}b. 
We have also compared to the case when the system starts from
equilibrium\footnote{Note that, as mentioned above, for such type of
initial conditions the dissipative corrections initially grow very
rapidly, which is not a physically plausible scenario~\cite{DMN}.}
($\Pi(t_0)=0$), see the curves $\Phi_{3(RK)}$ and $\Phi_{3}$. 
In both cases the SHASTA result is very good. 
There are some small deviations, however, this is due to the difference 
between the calculated and analytical expansion rates.  
We have checked our calculations using the exact expansion rate $\theta = 1/\tau$
and in this case not only the 2nd-order calculations but also the 1st-order ones 
are smooth and very accurately match the Runge-Kutta results.

\section{Summary and Outlook}

In this work we have focused on testing numerical solutions of first and
second order theories of relativistic hydrodynamics with bulk viscosity
using the SHASTA flux corrected transport algorithm. This is a rather
efficient and fast algorithm for solving causal fluid transport on a fixed
grid; it provides accurate solutions of ideal hydrodynamics with
minimal numerical viscosity and prediffusion and can be easily adapted
to multi-dimensional problems. In fact, the algorithm can also be used
to solve the relaxation equations of the 2nd-order approach
simultaneously with the conservation equations without resorting to
other numerical schemes, which may reduce the computational time and
complicate the problem and its implementation further.

The 1st-order theory of viscous hydrodynamics provides the proper
description of long-wavelength, low-frequency density waves in a
fluid. The 2nd-order theory introduces relaxation equations for the
dissipative fluxes thereby maintaining causality. Its solutions
converge to those of the 1st-order theory over time scales larger
than the relaxation times. It has been argued that these relaxation
times might be on the order of the microscopic time scales in the
problem and that the 2nd-order theory is therefore no better
approximation to the dynamics than the 1-st order approach.

Our numerical solutions with the SHASTA algorithm, however, indicate
that the accuracy and stability of the solutions of the 2nd-order
theory is significantly better than in the 1st-order theory, even if
the calculated local expansion rate is smoothed over a few fluid
cells: the solution of the Riemann wave with viscosity in the
1st-order approach produces oscillations which are absent from the
2nd-order theory. 
This observation holds for virtually any amount of dissipation. 
Also, the numerical problems encountered in the 1st-order
approach get milder if the speed of sound is smaller (which reduces
the acceleration of the flow) but worse if the equation of state is stiff.

These observations are valid in 2 or 3 dimensional cases \cite{Molnar2009}, therefore in conclusion, 
we believe that for general purpose codes the 2nd order theory 
is not only more general, but also more stable and reliable even numerically.
Although using the 2nd-order theory it is computationally more intensive since
the dissipative quantities have to be propagated in time, its implementation
into some existing numerical codes which solve hyperbolic partial differential equations 
in conservation form, does not require more effort than adding the 1st-order corrections.

Regardless on the difference between numerical methods, using the same initial 
conditions and corresponding physical quantities, all numerical results should be 
very closely the same.
Unfortunately, in case of dissipative fluid dynamics there are only a few simple 
solutions where the numerical accuracy can be tested in great detail.
However, taking into account that actual applications of relativistic fluid dynamics in 
modeling relativistic heavy-ion collisions need several other crucial approximations, 
(introducing additional uncertainties and parameters linked to the fluid dynamical 
calculation, such as initial conditions and freeze-out),  
it is of great importance that the numerical fluid dynamical methods should be very carefully 
investigated, tested and documented in various situations.
To our knowledge, in most publications this topic is rather forgotten and/or undisclosed.    
The other reliable possibility and recommendation would be to check the fluid dynamical codes 
against kinetic theory  \cite{Molnar:2008xj,Huovinen:2008te}, which on the other hand  would also 
'validate' transport codes in the fluid dynamical regime.

Note added: During the preparation of this manuscript we became aware of
the very recent work by the Brazilian group \cite{Denicol:2008rj,Denicol:2008ua},
on the shock propagation and stability in causal 1+1 dimensional dissipative hydrodynamics,
using the smoothed particle hydrodynamics (SPH).
This important problem was also investigated by us with the relaxation method presented in this paper, 
and lead to very promising agreement with kinetic theory \cite{Bouras:2009nn}.


\section*{ACKNOWLEDGMENTS}

The author thanks A. Dumitru for the guidance, discussions, careful reading and corrections.
Furthermore, the author is thankful to D. H. Rischke for the original version of ideal
SHASTA code and many important comments and suggestions.
The author also thanks to H. Niemi for checking independently some results presented
and for his clarifying comments and recommendations.
Enlightening discussion on the topic with T. S. Bir\'o, L. P. Csernai, M. Gyulassy, 
J. A. Maruhn and I. Mishustin are noted.
Constructive comments by an anonymous referee are warmly noted. 
E.~M., gratefully acknowledges support by the Alexander von Humboldt foundation.


\section*{APPENDIX A}

Here we approximately extract and quantify the numerical errors and uncertainties as a function 
of time and viscosity.  
Since the exact solution for the Riemann problem with dissipation is unknown, 
the deviation from analytic solutions cannot be measured. 
However, since the 2nd order solutions converge to the 1st-order solutions at late times, 
we may quantify the error and deviations between the outcomes.

\begin{figure}[!htb]
\centering
\includegraphics[width=7.5cm, height = 5.5cm]{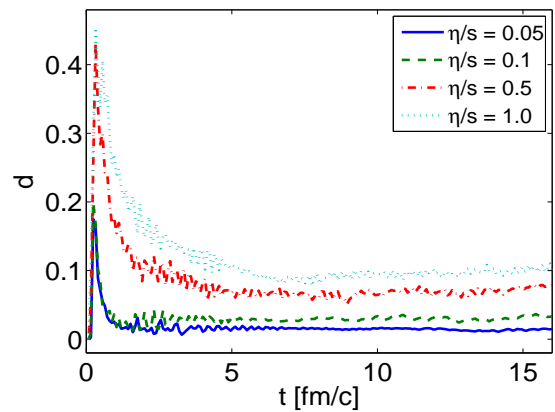}
\caption{The relative difference between the 1st-order and 2nd-order theory as a function 
of time, measured by the above integral, for values of viscosity given in the figure.}
\label{fig15}
\end{figure}

Therefore, we introduce the following relative deviation (point by point) between the flow velocities 
of matter and measure it by the following integral, 
\bea\label{deviation}
d(t) = \frac{1}{N}\int^{1.5}_{-1} d\xi \, \left|v_{2nd}(\xi)  - v_{1st}(\xi) \right|\, ,
\eea
where the normalization factor, $N=\int^{1.5}_{-1} d\xi \left|v_{2nd}(\xi) \right|$.
Note that, the 1st-order and 2nd-order theories start from different initial values, 
hence the initial deviation.
We also observe that increasing the viscosity the results start to diverge at late times, 
signaling the instability and large errors in the numerical solution of the 1st-order theory.
This was apparent in all figures in Sect. \ref{sl_diss}.

\begin{figure}[!htb]
\centering
\includegraphics[width=7.5cm, height = 8.5cm]{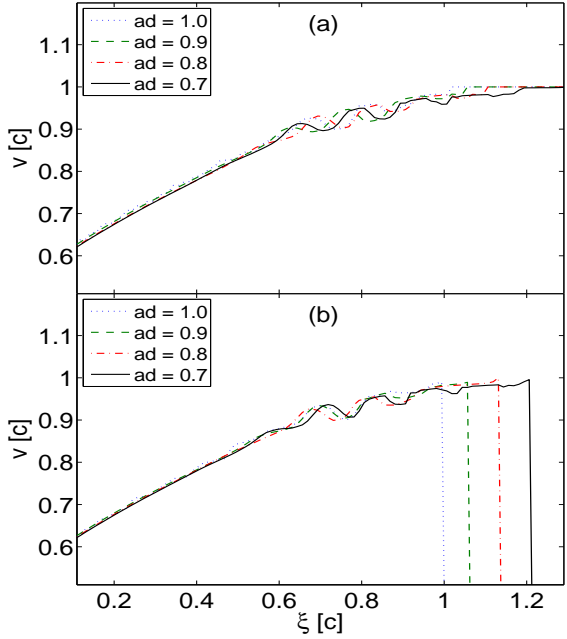}
\caption{
The velocity profiles as calculated numerically at $t = 16$~fm/c for a dissipative fluid with $\zeta/s = 0.1$ in
1-st order theory using different antidiffusion mask coefficients (ad) as shown on the figure.
The initial condition in subplot (a) is $v = 1$ for $z>0$, guarantees a continuous solution at the 
boundary to the vacuum. 
In (b) we have $v = 0$ for $z>0$.}
\label{fig16}
\end{figure}

Next we discuss and show further examples and test results using different initial
conditions and antidiffusion mask coefficients.
Using the purely conventional initial value for the velocity of vacuum,
$v = 1$ for $z>0$, we make sure that the fluid dynamical solution is continuous
at the boundary to vacuum.

This specific test ensures that the oscillations in the 1-st order theory are not caused
by the sharp boundary and that the oscillations propagate outwards and
not inwards from the vacuum, see Fig. \ref{fig16}a.
Therefore, this effect is mainly due to two things.
First the higher order derivatives of the flow velocity appear in the equations and therefore
even very small fluctuations in the flow field are enhanced and couple back into the solution.
The second is a purely numerical problem which unfortunately affects of the numerical
scheme and its accuracy since the SHASTA algorithm was not explicitly designed
to solve the relativistic Navier-Stokes equations.

\begin{figure}[!htb]
\centering
\includegraphics[width=7.5cm, height = 8.5cm]{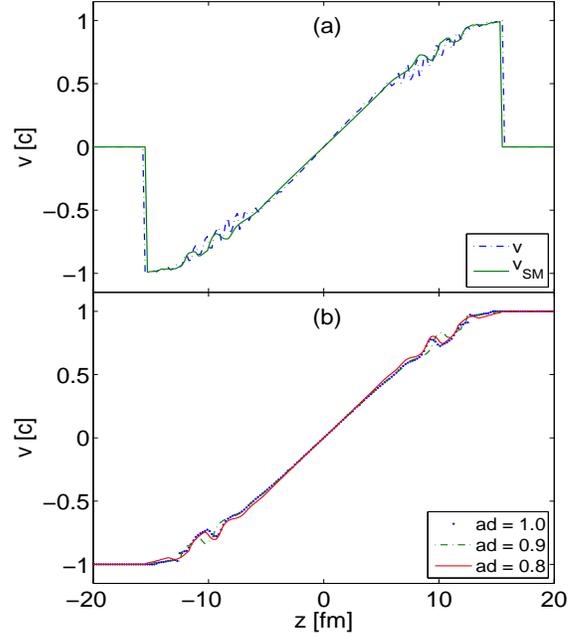}
\caption{
The velocity profile for a Bjorken expansion in 1st-order theory for $\zeta/s = 0.2$
after $\Delta t = 8$~fm/c evolution.
On (a) the un-smoothed vs. the smoothed $v_{SM}$ velocity profile with ad=0.8 mask coefficient.
The continuous boundary condition for the velocity with various mask coefficients on (b).}
\label{fig17}
\end{figure}

We also see the effect of a further numerical artifact namely reducing the antidiffusion
coefficient by $10\%,20\%$ and $30\%$ not only increases the entropy in the system,
but also gives result to non-linear changes and differences in the solutions.
The standard version of the algorithm uses a mask coefficient equal to 1.
We can see in Figs. \ref{fig16} and \ref{fig17}b that using a $20\%$ smaller mask is
reasonably close to the solution with the standard value of the mask, but more importantly
the results become much smoother.

In Fig. \ref{fig17}a we have plotted the velocity profile calculated with a $20\%$ reduced antidiffusion
for a smoothed and un-smoothed expansion rate.
We can see that the smoothing affects the solution positively leading to even less prediffusion into 
vacuum in this particular case.

\section*{APPENDIX B}

Here we recall the solutions for the expansion into vacuum
in case of a perfect fluid following \cite{Centrella:1983cj}.
Introducing the similarity variable, $\xi = z/t$, the partial derivatives
transform as, $\partial_t = -  (\xi/t) \, (d/d\xi)$ and
$\partial_z = (1/t) \, (d/d\xi)$.
Therefore the equation for the energy and momentum in terms of the rest frame
quantities becomes,
\bea
&&(v - \xi) \gamma^2 \frac{d e}{d\xi}
+ \left[v + (v - \xi) v^2 \gamma^2 \right]\frac{d P}{d\xi} \\ \nonumber
&+& \gamma^2(e + P)\left[(v-\xi) (2v\gamma^2) + 1\right]\frac{dv}{d\xi} = 0 \, ,\\
&&(v - \xi)v \gamma^2 \frac{d e}{d\xi} + \left[1 + (v - \xi) v \gamma^2 \right]\frac{d P}{d\xi} \\ \nonumber
&+& \gamma^2(e + P) \left[(v - \xi)(1 + 2v^2\gamma^2) + v \right]\frac{dv}{d\xi} = 0 \, .
\eea
Using the standard EOS, $P = c^2_s e$, the vanishing determinant of the above system of equations
leads to the expression for the characteristic variable, $\xi = \frac{v \pm c_s}{1 \pm v c_s}$.
The correct solutions implies eq. (\ref{simvariable}), hence we are lead to the
following trivial equation,
\bea
\frac{de}{e} = -\frac{(1 + c^2_s)}{c_s} \frac{dv}{(1 - v^2)} \, ,
\eea
which with the corresponding initial conditions given in sect. \ref{riemann} leads to
the results presented before.

Viscosity is introduced as, $P = c^2_s e + \Pi$, where the expansion rate is,
\bea
\partial_{\mu}u^{\mu} &=&  \left(1 - v \xi \right)\frac{\gamma^3}{t}  \frac{dv}{d\xi}\, .
\eea
Using the expression for the dissipative pressure in 1st-order theory, $dP/d\xi$, leads to
terms containing, $d^2v/d\xi^2$, $(dv/d\xi)^2$, $d\zeta/d\xi$ and
$dt^{-1}/d\xi = - t/\xi$, therefore even in this simple case the exact solution is unknown.
In 2nd-order theory the relaxation equation is,
\bea
(v - \xi) \gamma \frac{d \Pi}{d\xi} - \frac{\zeta }{\tau_{\Pi}} (1 - v\xi) \gamma^3 \frac{dv}{d\xi}
= - \frac{t}{\tau_{\Pi}}\, \Pi  \, ,
\eea
while the derivative of the pressure reduces to, $dP/d\xi = c^2_s de/d\xi + d\Pi/d\xi$.
In conclusion we see that for dissipative fluids the equations depend explicitly on the time in
the local rest frame and the similarity of the flow is broken.



\end{document}